\documentstyle[prl,twocolumn,aps,epsf]{revtex}
\begin{document}

\title {Charged particle fluctuations and microscopic models of 
nuclear collisions}
\author{Q. H. Zhang$^{a}$,  V. Topor Pop$^{a}$, S. Jeon$^{a,b}$  and C. Gale$^{a}$ }
\address{ $^a$Physics Department, McGill University, Montreal QC H3A 2T8, 
Canada\\
       $^b$RIKEN-BNL, Research Center, BNL, Upton, NY11973, USA}
\maketitle

\begin{abstract}
We study the event-by-event fluctuations of the charged particles and 
compare the results of different Monte-Carlo Generators (MCG): VNIb, HIJING, 
HIJING/B${\bar B}$ and RQMD.  
We find that the $D$-measure can be used to distinguish between the 
different gluon populations that are present in the MCG models.
On the other hand,  the value of the $D$-measure
shows high sensitivity to the rescattering effects in VNIb model,
but lower sensitivity to the rescattering effects 
 in RQMD  model.  We also find that the $D$-measures from AA are  
consistent with the $D$-measures from $pp$ 
for all generators except VNIb. 
Therefore, any deviation among the values of $D$-measure for different 
impact parameters and between  
 $pp$ and $AA$ collisions may indicate that either the rescattering effects 
play a key role in the interactions or there is 
new physics in $AA$ collisions. 

\end{abstract}

PACS number(s): 25.70 -q, 25.70 Pq,25.70 Gh

\section{Introduction}

One of the main purposes of high energy heavy-ion collisions 
is to produce a macroscopic size of  quark-gluon-plasma(QGP)\cite{Wong1,Hwa1}. 
With hadrons as final observables, 
several signatures have been suggested: 
$J/\psi$ suppression\cite{qm01}, single 
event fluctuations measurements\cite{BH99,HJ00,JK99,HY01,GM92,BR00,Z00}, 
 and void and gap searches\cite{HZ00}. 
It was also proposed in Ref.\cite{JK00,AHM00} that the quantity 
\begin{equation}
D(\Delta y)=\langle N_{ch}\rangle_{\Delta y} \langle \delta R^2\rangle_{\Delta y} \sim 
4\frac{\langle \delta Q^2\rangle_{\Delta y}}{\langle N_{ch}\rangle_{\Delta y}} ,
\label{xe1}
\end{equation}
 be used as a signature of QGP. Here $N_{ch}=N_{+}+N_{-}$
is the total number of charged particles, $R=N_{+}/N_{-}$ is the ratio between positive 
charge and negative charge and $Q=N_+-N_{-}$ is the net charge. The second 
moments of $R$ and $Q$ are defined as 
\begin{equation}
\langle \delta x^2\rangle =\langle x^2\rangle-\langle x\rangle^2,
\end{equation}
where $\langle ...\rangle$ means the 
average taken over all events and $\Delta y$ is the rapidity 
window in which we 
calculate the above quantities. The last step in Eq.~(\ref{xe1})
 is correct to leading order in $1/\langle N_{ch}\rangle $ and 
in the fluctuations. It is the observable defined by this last term 
of Eq.~(\ref{xe1})  which we calculate throughout this paper. 
It has been found that for a pion gas, 
the  $\frac{4\langle \delta Q^2\rangle}{\langle N_{ch}\rangle} $ is 
around 4 and 
for a QGP gas it is 
approximately $1$\cite{JK00,AHM00}; 
therefore, the $D$-measure has been proposed as a 
signature of QGP\cite{JK00,AHM00}. 

However, this observable ($D$-measure) has some caveats which have been 
discussed recently. Gazdzicki and Mrowczynski\cite{GM00}
have argued that $\langle \delta Q^2\rangle$ in $AA$ 
collisions could be determined by the  
number of participating protons. So the smallness of the $D$-measure 
may be just an indication of the smallness of 
$\frac{\langle N_{part}\rangle}{\langle N_{ch}\rangle}$.
Fialkowiski and Wit\cite{FW00} have commented that the $D$-measure is 
a rapidity dependent quantity and 
the prediction of the $D$-measure from PYTHIA/JETSET model is 
smaller than the estimated value for hadron gas and 
become much smaller (even less than 
the estimated value for QGP) when the rapidity region is very large.
 
In reply to the second criticism, Bleicher, Koch and one 
of us\cite{BJK00} argued  that the dependence of $D$-measure on 
rapidity is due to the 
following two facts: (1) In Ref.\cite{JK00}, 
the approximation 
$\langle N_{+}\rangle_{\Delta y}=\langle N_{-}\rangle_{\Delta y}$ was used 
which may be not 
fulfilled for heavy-ion collisions, so one needs to apply the 
correction $(\frac{\langle N_{+}\rangle_{\Delta y}}{\langle N_{-}\rangle_{\Delta y}})^2$.
(2) Also, it was assumed that the charge ratio 
fluctuated independently in each rapidity window\cite{JK00}.  
This is inappropriate 
due to global charge conservation, 
 and this brings up another correction factor 
$1-\frac{\langle N_{ch}\rangle_{\Delta y}}{\langle N_{ch}\rangle_{total}}$.
 After these 
corrections, it was found that the corrected measure $D_{corr}(\Delta y)$
\begin{equation}
D_{corr}(\Delta y)=\frac{D(\Delta y)}
{(\frac{\langle N_+\rangle_{\Delta y}} {\langle N_{-}\rangle_{\Delta y}})^2
(1-\frac{\langle N_{ch}\rangle_{\Delta y}}{\langle N_{ch}\rangle_{total}} )},
\label{e2}
\end{equation}
 predicted by UrQMD\cite{UrQMD} is around $2.5\sim 3.1$\cite{BJK00} which is 
consistent with the estimated value of $D$-measure 
for resonance gas\cite{JK99,HJ00}. 
In Ref.\cite{BJK00}, no significant differences have been found 
for $D$-measure values at SPS and RHIC energies, meaning that 
the $D$-measure has little energy dependence\cite{AHM00,BJK00}.  

As has been discussed, for example, in Open Standard Code and Routine (OSCAR) 
conferences\cite{OSCAR}, all  
MCGs\cite{VNIb,ARC,RQMD,VENUS,HIJING,HIJINGbb,ST,ZPC} used now are not simple 
codes, they contain different physical 
ingredients and assumptions; 
therefore, it is very interesting to study and compare theoretical predictions
from some MCGs, which are based on  different physics pictures.
In this paper we calculate the $D$-measure using the 
VNIb\cite{VNIb}, RQMD v2.4 \cite{RQMD}, 
HIJING v1.35 \cite{HIJING}, and 
HIJING$/B{\bar B}$ v1.10 \cite{HIJINGbb} models (See section II for a 
short discussion of these models).  
One of the striking results is that the values of $D$-measure from 
VNIb model (running with rescattering turned off) is much 
less than the values of $D$-measure 
from RQMD, HIJING, HIJING/$B{\bar B}$ and UrQMD\cite{BJK00} models. 
The reason for this difference could be 
the different number of gluons embedded in the model.
In heavy-ion collision processes, if 
the degrees of freedom are partons or hadrons at the initial 
stage of collisions, 
 we will expect to have a different charged fluctuation 
if the rescattering effects do not play a key role in interactions.
{\em In this sense, the 
 $D$-measure could be a signature of QGP.} 
However, to be considered as a {\em good signature of QGP}, 
$D$-measure values must be compared also between  
nucleus nucleus (AA) and proton-proton (pp) collisions.
If the $D$-measure is dominated by the physics just before hadronization,
any differences between the values obtained from AA and pp 
collisions indicate that either rescattering effects are strong, 
or a signature of new physics (e.g. presumably QGP) in AA collisions.  

One of the aims of this work is to perform a systematic 
study of the charged fluctuations using many 
of the available and popular event generators. We 
believe this exercise is valuable first 
to establish the fluctuations as a robust variable, then to 
interpret the physical information the measurements contain. We 
investigate the effects of rescattering and we also consider the 
impact parameter dependence of the signal.

This paper is arranged in the following way: In Sec. II, 
using VNIb, RQMD v2.4, HIJING v1.35, HIJING/$B{\bar B}$ v1.10 models 
we calculate the $D$-measure for 
 AA  collisions at total centre of mass (c.m) energy 
 $\sqrt{s}=200$A GeV and we find that the $D$-measure of 
VNIb (with rescattering turned off) is much less than the 
$D$-measure of other models and an explanation is given. In Sec. III, we 
study the rescattering effects on VNIb and RQMD. Our results show that 
rescattering effects may spoil the signature of physics in 
VNIb; on the other hand, the rescattering effects on the $D$-measure 
of RQMD are less dramatical.  A comparison of $D$-measure 
between pp and AA are performed and the similar value between 
of $D$-measure between $pp$ and $AA$ are explained within a ``participant 
model''. 
Finally, our discussions and conclusions are given in Sec. IV.

\section{ The $D$-measure with different MCGs}

As in Ref.\cite{BJK00},
we will calculate the $D$-measure using 
VNIb\cite{VNIb}, RQMD v2.4 \cite{RQMD}, 
HIJING v1.35 \cite{HIJING}, HIJING$/B{\bar B}$ v1.10 \cite{HIJINGbb} models.   
In the following we briefly outline the main features of those models. 

In HIJING \cite{HIJING}, the physics of minijets is addressed explicitly in
perturbative Quantum ChromoDynamics (pQCD). The cross sections for hard
parton scattering are calculated at the leading order and a K-factor is
invoked to account for higher-order corrections. Soft contributions are
modeled by diquark-quark strings with gluon kinks induced by soft gluon
radiation. Jet quenching and shadowing can also be treated in this 
approach. HIJING$/B{\bar B}$ \cite{HIJINGbb}  is based on HIJING and a
baryon junction mechanism is introduced in order to understand the
longitudinal distributions of anti-baryons from pA and AA collisions at SPS
energies. The junction-antijunction loops
that arise naturally in Regge phenomenology are also included  
in the calculation.  Final state interactions among produced hadrons are 
implemented neither in HIJING nor in HIJING$/B{\bar B}$.
RQMD \cite{RQMD} is a transport approach for hadrons and resonances, with
initial-state hadronic string generation. There, 
overlapping strings may fuse into colour-ropes.
The fragmentation products 
from ropes, strings, and resonances may then interact with each other 
and with the original nucleons. In this model copious 
rescatterings lead to the development of collective flow 
and can drive the system towards local equilibrium.

As opposed to RQMD, HIJING and HIJING$B{\bar B}$, VNIb treats  
a nuclear collisions in terms of parton-parton interactions. It 
uses a transport algorithm to follow the evolution of the many-body 
system of interacting partons and hadrons in phase space. 
For hadronization, 
VNIb uses a parton-cluster  formation and fragmentation 
approach. Rescattering among partons and hadrons 
is included in the code. One important feature of  
VNIb is that,  at RHIC energy, it generates a
substantial gluon population.  Those then play an important role in 
the simulation of RHIC data in VNIb. 

\begin{figure}[h]\epsfxsize=8cm
\centerline{\epsfbox{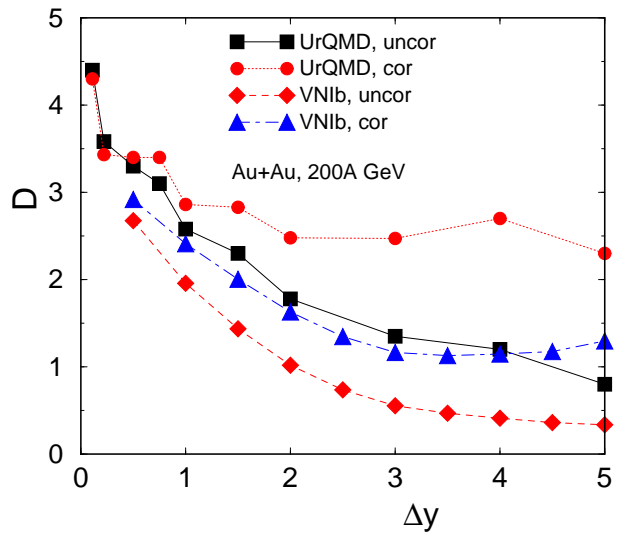}}
\caption{ $D$-measure vs. rapidity, $y_{cm} \pm \frac{\Delta y}{2}$
for Au+Au central collisions ($b \le 2$ fm) at  
total c.m energy $\sqrt{s}=$ 200A GeV.
Full squares and diamonds denote the results 
predicted by UrQMD (from Ref. [16]) and 
VNIb respectively. Circles and triangles 
denote the results obtained when taking into account the correction factors
(see text for explanations).}
\label{fig4}
\end{figure}
In Fig.~\ref{fig4}, the values 
of the $D$-measure from VNIb 
(rescattering turned off) vs. the rapidity window are shown. 
For comparison, the results from UrQMD\cite{BJK00} are also included in 
the plot. 

We notice that there are big differences between 
the values of the $D$-measure from VNIb and those from UrQMD. 
 Applying the correction 
method given in Ref.\cite{BJK00}, we calculate also the 
corrected values of $D_{corr}$ (see Eq.~(\ref{e2})) and we obtain   
a higher value for a large rapidity window.  
  For a rapidity window around 
$(-2,2)$ we find that  the value of $D_{corr}$ is around one. 
For smaller rapidity window, the value of $D_{corr}-$measure is bigger than  
one, and this can be explained by the 
fact that small windows will not catch all the decay  
products of a resonance.
If we analyze the correction factor, 
$1-\frac{\langle N_{ch}\rangle_{\Delta y}} {\langle N_{ch}\rangle}$ 
given in Ref.\cite{BJK00}, we find that 
for the whole kinematic phase space this correction factor should 
be zero and can not be used for very 
larger rapidities, so  we must overlook the results  
for larger rapidity windows ($\Delta y > 4$) in Fig.~\ref{fig4}.  

\begin{figure}[ht]\epsfxsize=8cm
\centerline{\epsfbox{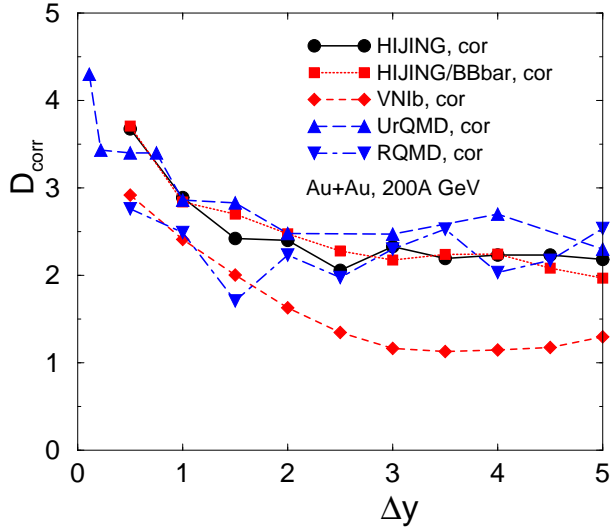}}
\caption{ The corrected values of $D_{corr}$-measure from  
VNIb, UrQMD, RQMD, HIJING,HIJING/$B{\bar B}$ vs. rapidity  
$y_{cm} \pm \frac{\Delta y}{2}$ for central $Au+Au$ 
collisions ($b\le 2fm$) at total c.m. energy $\sqrt{s}=200A\,\, GeV$.}
\label{fig5}
\end{figure}

The corrected values $D_{corr}$
obtained from the predictions of VNIb (rescattering turned off), 
RQMD, HIJING, HIJING/$B{\bar B}$ models are shown in Fig.\ref{fig5}.  
The values from VNIb are 
lower than the values predicted by other  MCG models.
The main difference can be due to 
the different number of gluons embedded
in VNIb, which is higher than in any other MCGs considered 
here.

 The predictions 
obtained from all the above models, except VNIb, are 
consistent with each other in the limit of statistical errors.
If rescattering among produced hadrons is not 
a dominant effect during heavy-ion collisions, then 
the $D$-measure should be determined by the 
physics just before hadronization as assumed in 
Ref.\cite{JK00,AHM00}.  
According to this picture, 
string model codes, like UrQMD, RQMD, HIJING, HIJING$/B{\bar B}$ 
(we note that UrQMD and RQMD include also hadronic picture in the
code),  form strings using 
the quarks or diquarks from two collided  nucleons and there is no, or very few 
gluons\cite{Hijing}. So those quarks and antiquarks 
will dominate the final state charge fluctuations. On the other hand, 
for a model like VNIb which contains a large population 
of gluons, the observed $D$-measure should be different from 
the results calculated from RQMD and HIJING. It is known that 
if there are only gluons in the initial state of heavy-ion 
collisions and if we consider gluon 
fusion processes (like $gg\rightarrow q{\bar q}$),
then  the charge fluctuation in a larger rapidity window (for our case 
from $-2$ to $2$ for example)
\begin{equation}
\langle \delta Q^2\rangle \sim 0,
\end{equation}
as the charge is almost conserved in that window;
 those gluons also  produce large number of 
charged particles.  
Thus the $D$-measure for a gluon gas should be very small. In the 
VNIb code, we have quarks, antiquarks and gluons. 
  By examining the parton population in VNIb, we find that the 
ratio of the number of gluons to 
the number of quarks and antiquarks 
 from the runs for $Au+Au$ collisions at 
$200~~GeV$ is around $1.2$. 
If we exclude the extra 
valence quarks (those valence quarks will mainly 
contribute to fragmentation regions) coming from nucleons (so that 
$\langle N_q\rangle =\langle N_{\bar q} \rangle$ ), then the ratio 
is $1.8$.  That is, the central rapidity region is the most gluon dominated 
region in VNIb code\cite{GK93}.  This could explain why the $D$-measure 
from VNIb is less than the $D$-measure from RQMD and HIJING.

This  analysis shows that one obtains different 
values of $D$-measure owing to the different physics embedded in the MCG; 
however, to draw any final 
conclusion, we should have new theoretical predictions using 
models such as, for example, ARC\cite{ARC} and 
  compare them with predictions from VNIb and ZPC\cite{ZPC}.
ARC is based on hadronic physics and pictures nuclear collisions 
in terms of nucleon-nucleon collisions. For nucleon-nucleon collisions,
the model uses data from experiment. As opposed to 
RQMD and HIJING, there is no string picture in ARC.  Because of this we 
expect that 
ARC should give a value of the $D$-measure around 
three. 
On the other hand, ZPC\cite{ZPC} is a versatile simulation program that can
use initial parton distributions from any source as input, and can study 
parton evolution and rescattering.  
However, there is  no hadronization algorithm implemented 
in the code. One could use the parton mode of ZPC to calculate directly the $D$-measure 
which should be less than one, following the reasoning in \cite{JK00,AHM00}.

Finally, we mention that 
one can account for all final state particles in a MCG model,
which is  not the case  in heavy-ion experiments because of 
 the fact that detectors can not detect 
all charged particles. 
So, we can imagine that there is no charge conservation among the 
 {\it detected } particles. Here,  we will 
discuss detector efficiency for two cases: 
{\em Case I}: if we assume that the 
detector efficiency is the same for both positive and negative 
particles in each event,
then the $\langle R^2\rangle$ and $\langle R\rangle$  should remain 
the same.  We notice that as the measured charged particles 
$f \langle N_{ch} \rangle$ becomes smaller ( here $f$ is the 
detector efficiency which represents the ratio of 
the measured particles to the produced particles, 
$\langle N_{ch}\rangle$ is the production particles),  the $D$-measure
 will become smaller too. {\em Case II:} We assume two 
Poissonian distributions for both produced positive and negative 
charged particles, that is 
\begin{equation}
P(N_i)=\frac{\langle N_i\rangle^{N_{i}}}{N_i!}\exp(-\langle N_i\rangle)
~~~~i=\pm.
\end{equation} 
We further assume that due to the detector efficiency, 
the observed particle number $S_{i}$ 
 follows a Binomial distribution ($S_{\pm}\le N_{\pm}$)
\begin{equation}
P(S_i|N_i)=\frac{N_i!}{S_i!(N_i-S_i)!}f^{S_i}(1-f)^{N_i-S_i}~~~i=\pm.
\end{equation}
Then one can easily verify that the observed charged particles 
have again a Poisson distribution
\begin{eqnarray}
P(S_i)&=&\sum_{N_i=S_i}^{\infty}P(N_i)P(S_i|N_i)
\nonumber\\
&=&\frac{(\langle N_i\rangle f)^{S_i}}{S_i!}\exp(-\langle N_i\rangle f)
~~~i=\pm.
\end{eqnarray}
From above we have 
\begin{equation}
\langle \delta Q^2\rangle=f\langle N_+\rangle
+f\langle N_-\rangle 
-2f^2\langle \delta N_+ \delta N_{-}\rangle.
\end{equation}
Thus
\begin{equation}
D=4-8f\frac{\langle \delta N_{+}\delta N_{-}\rangle}
{\langle N_+\rangle+\langle N_{-}\rangle}.
\label{exxf}
\end{equation}
This indicates that when $f$ becomes smaller then 
the $D$-measure will become bigger. This is different from the 
conclusion in the Case I.  In Case I, there is strong correlation 
between the detector efficiencies of positive and negative 
charge particles in each event; on the other hand, 
there is no correlation between detector efficiencies 
of  positive particles and negative charge particles in Case II.
  The practical case can be more complex. However, 
 as shown here that $D$-measure is sensitive to the detector 
efficiency and we need to exercise caution when comparing theoretical 
predictions with data. 

\section{Rescattering effects on the $D$-measure}

\subsection{Rescattering effects on the $D$-measures of VNIb and RQMD}

Large  rescattering effects can destroy 
the physical correlations which originate from the QGP phase. 
Then we will only get a hadronic resonance gas  signature, 
$D \sim 3$\cite{SS00}. Rescattering effects 
depends on two factors, one is the time that particles need to 
go through the collision region, another one is the density 
in the collision region. Those two effects will 
determine the mean free path of particles in the interaction region. 
For high energy collisions, the time that particles 
needed to  pass through the collision regions is short, since  
the density is higher.  No simple 
relation exist to determine the effects of rescattering 
on the $D$-measure yet.

 We note that $D$-measure values from UrQMD model have no  
impact parameter dependence up to very peripheral collisions and we know that 
if the impact parameter of AA collisions is very large the nuclei-nuclei 
collision will be only a superposition of pp collisions
(may be one or several pp collisions) and rescattering 
effects will become smaller. On the other hand, if the impact parameter is 
smaller, then rescattering effects could play an important role. 
Most MCGs use the following scheme
\vskip 0.4cm
\begin{eqnarray}
A+A&=&\sum (nucleon + nucleon)     +
\nonumber\\
&&
\nonumber\\
&& (secondary~~ particle + secondary~~ particle)
\nonumber\\
&&
\nonumber\\
&&+(secondary~~ particle +nucleon).
\label{e4}
\end{eqnarray} 
\vskip 0.4cm
UrQMD model predictions show that $D$-measure is almost impact parameter 
independent, and this indicates that rescattering 
effects do not play a key role for the values of 
$D$-measure at RHIC energy\cite{BJK00}. 
 The physics should then be dominated 
by the simple $nn$ collisions if the model employed the 
scheme described by Eq.~(\ref{e4}).
We also remark that the predictions from 
UrQMD at SPS energy are larger than the predictions at RHIC
energy. The main differences could perhaps be attributed to  
the mix of hadronic degrees of freedom and string degrees of freedom.
 When energy is higher the string formation dominate the collisions 
process,  while when energy is lower the hadronic picture does. 
This may explain why the values of $D$-measure at SPS energy are 
slightly higher than the values at full RHIC energy. 

\begin{figure}[ht]\epsfxsize=8cm
\centerline{\epsfbox{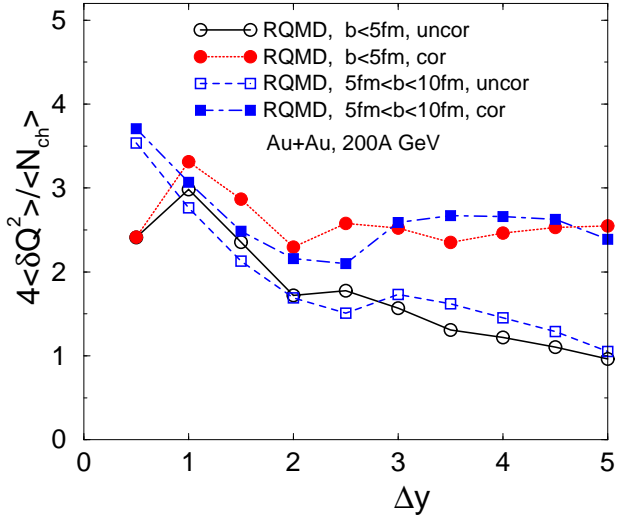}}  
\caption{$D\sim 4\langle \delta Q^2\rangle/\langle N_{ch}\rangle $  
values from RQMD vs. rapidity  $y_{cm} \pm \frac{\Delta y}{2}$
for Au+Au  collisions at  total c.m. energy $\sqrt{s}$=200A GeV.
The circles are the results for impact parameter range 
$b \le 5$ fm and the squares are the results for 
$5 \le b \le 10$ fm. The full and empty symbols are corrected and 
uncorrected values, respectively.}
\label{fig10}
\end{figure}

In Fig.~\ref{fig10}, we plot 
the values of $D$-measure from RQMD  
(rescattering turned on)  
in order to study the impact parameter dependence of rescattering,
for two different impact parameters regions. 
Analyzing the results from Fig. \ref{fig10}, we note that  
rescattering is slightly higher for central collisions 
($b< 5\,\, fm$) in comparison 
with peripheral ones ($5 < b < 10\,\, fm$) at low $\Delta y$.
Also, the results seems to indicate that 
rescattering effects 
are negligible for a rapidity window $\Delta y>1.0$.  

\begin{figure}[ht]\epsfxsize=8cm
\centerline{\epsfbox{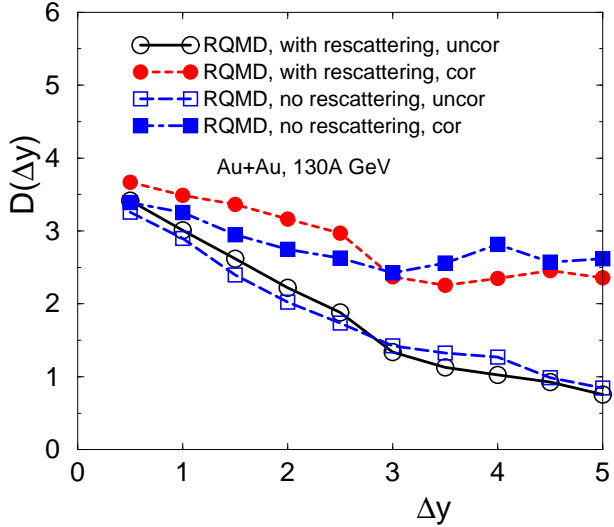}}  
\caption{$D(\Delta y)$ values from RQMD vs. 
rapidity  $y_{cm} \pm \frac{\Delta y}{2}$
for Au+Au  collisions at  total c.m. energy $\sqrt{s}=130\,\,$A GeV.
The circles are the results with rescattering turned on 
while the squares are the results for the case without 
rescattering. The full and empty symbols are the corrected and 
uncorrected values, respectively.}
\label{fig130}
\end{figure}

We also calculate 
the $D$-measure value from RQMD model at $130 \,\,$GeV with rescattering turned 
on and off. The results are shown in Fig.~\ref{fig130}. 
It is clear that rescattering effects on the 
value of $D$-measure are within $10\%$. This result is consistent with 
those in Fig.~\ref{fig10}.

\begin{figure}[ht]\epsfxsize=8cm
\centerline{\epsfbox{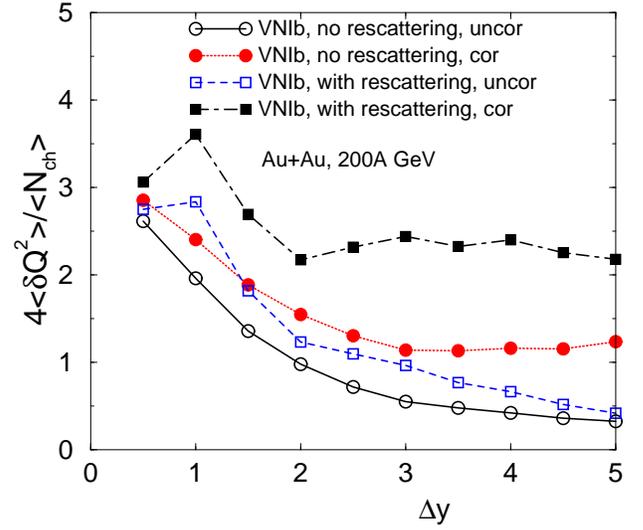}}  
\caption{$D\sim 4\langle \delta Q^2\rangle/\langle N_{ch}\rangle$  
values from VNIb vs. rapidity  $y_{cm} \pm \frac{\Delta y}{2}$
for Au+Au central collisions at total c.m. energy $\sqrt{s}$=200A GeV.
The circles corresponds to the run 
with rescattering turned off while the squares corresponds 
to the run with rescattering turned on. 
The full and empty symbols are corrected and 
uncorrected values, respectively.}
\label{figxx}
\end{figure}

In Fig.~\ref{figxx},  
the values of the $D$-measure from VNIb 
(rescattering turned on and off) are shown. 
It is found that the values of the $D$-measure 
are around $2.3$  for rescattering 
turned on and are smaller with
rescattering turned off ($\approx 1.0$).
These results show the different effects of rescattering in 
VNIb and RQMD v2.4 (see Fig.~\ref{fig10},~\ref{fig130}). 
Those could be related 
to the different densities of hadronic matter at the 
beginning stage of hadronization.  As the density of 
hadronic matter of VNIb is higher than the density of RQMD, 
 rescattering plays a more important role in 
VNIb.

From above we conclude that the values of the $D-$measure  
from the RQMD and UrQMD models have no impact parameter dependence.  
 Therefore we strongly suggest that the RHIC experiments must determine
the impact parameter dependence of  
$D$-measure to verify the above results.
If the experimental values indicate a
different trend in comparison with theoretical predictions,
we may consider  that the idea of Eq. (\ref{e4}) is too simple and  
one needs to involve other effects, 
such as the fact that  the parton distributions functions in nuclei 
are potentially different 
from the parton distributions functions in nucleon.

If $D$-measure for AA collisions is 
dominated by single nn interactions, one can imagine that  at lower energy, 
the single nn collisions is dominated by hadronic picture (cluster
picture), and at higher energy, nn collisions 
can see the content of nucleon. When energies increase,
it is expected that gluon should have also higher contribution. Based on the 
above assumption, if we plot the $D$-measure 
of $pp$ collision as the function of collision energy there should 
exist a drop from three to one. 
Even if  there is no such drop, one needs to get the trend that 
$D$-measure is really high at  lower energy and becomes smaller 
at high energy. 
Similar analyses should be performed for heavy-ion collisions too, 
in order to obtain energy dependence of 
$D$-measure from Bevalac to LHC energies. 

On the other hand, if the rescattering effects 
play a key role as in VNIb model, 
the signature of the initial stage of heavy-ion collisions 
will be lost.
However, combined analysis of $D$-measure with 
other 
signatures of QGP probably could still give us some more information about the 
unknown matter created in the early stages of AA collisions.

\subsection{$D$-measure for pp and AA}

We compare the values of $D$-measure 
for  $pp$ and $AA$ collisions obtained  
from VNIb (rescattering turned off), VNIb (rescattering turned on), 
HIJING v1.35, HIJING/$B{\bar B}$ v1.10, RQMD v2.4 
(rescattering turned on) in Fig.~\ref{fig12}(a-e).
 We note that the $D$-measure 
for AA collisions from VNIb without rescattering,
HIJING, HIJING/$B{\bar B}$ and RQMD  are all consistent with 
the $D$-measure for $pp$ interactions. 
On the other hand, the 
values of $D$-measure for AA from VNIb with rescattering 
are larger than the predictions for $pp$ 
due to rescattering effects.  

\vskip 0.7cm
\begin{figure}[ht]\epsfxsize=7.cm
\centerline{\epsfbox{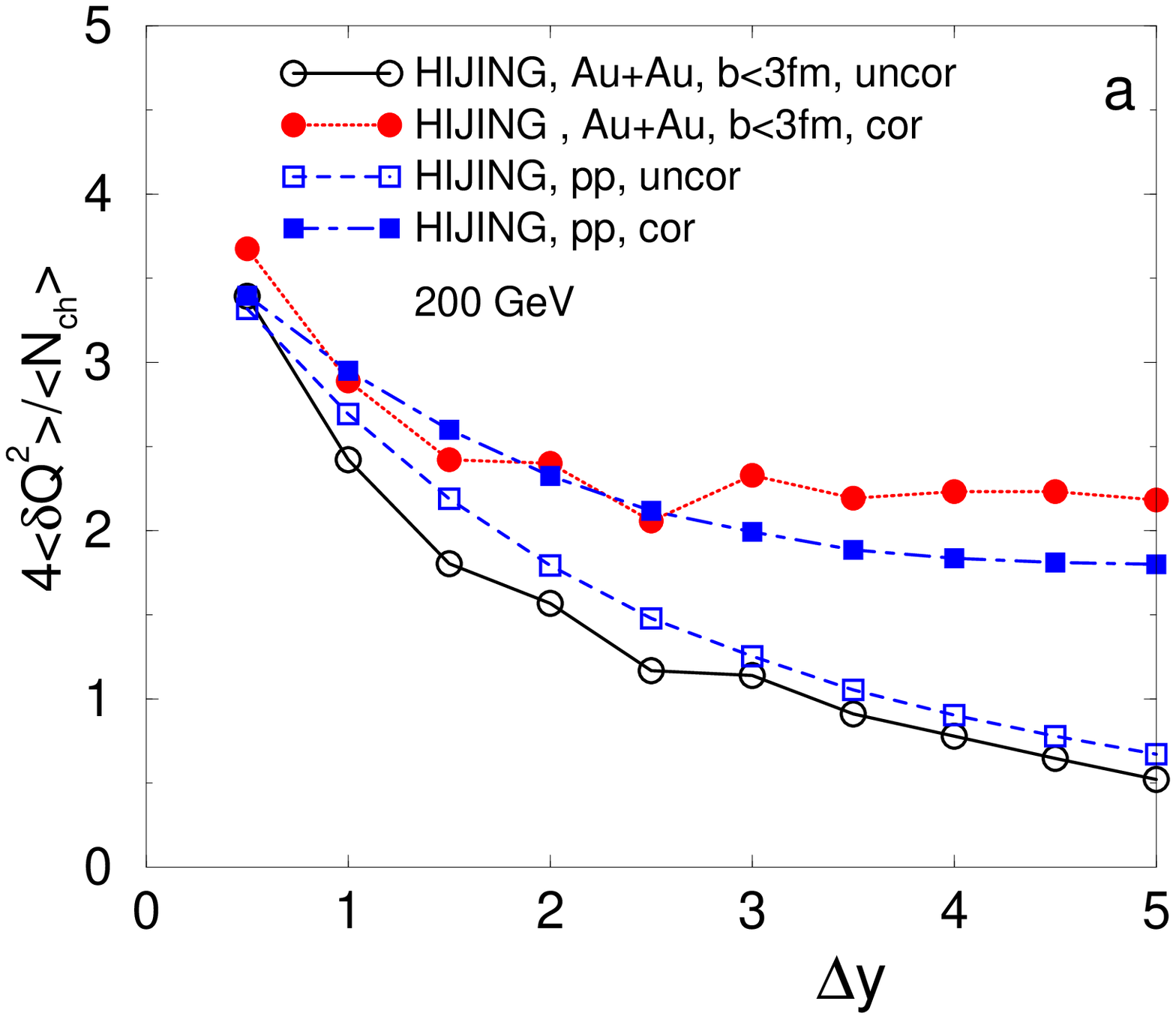}}
\end{figure}
\begin{figure}[ht]\epsfxsize=7.cm
\centerline{\epsfbox{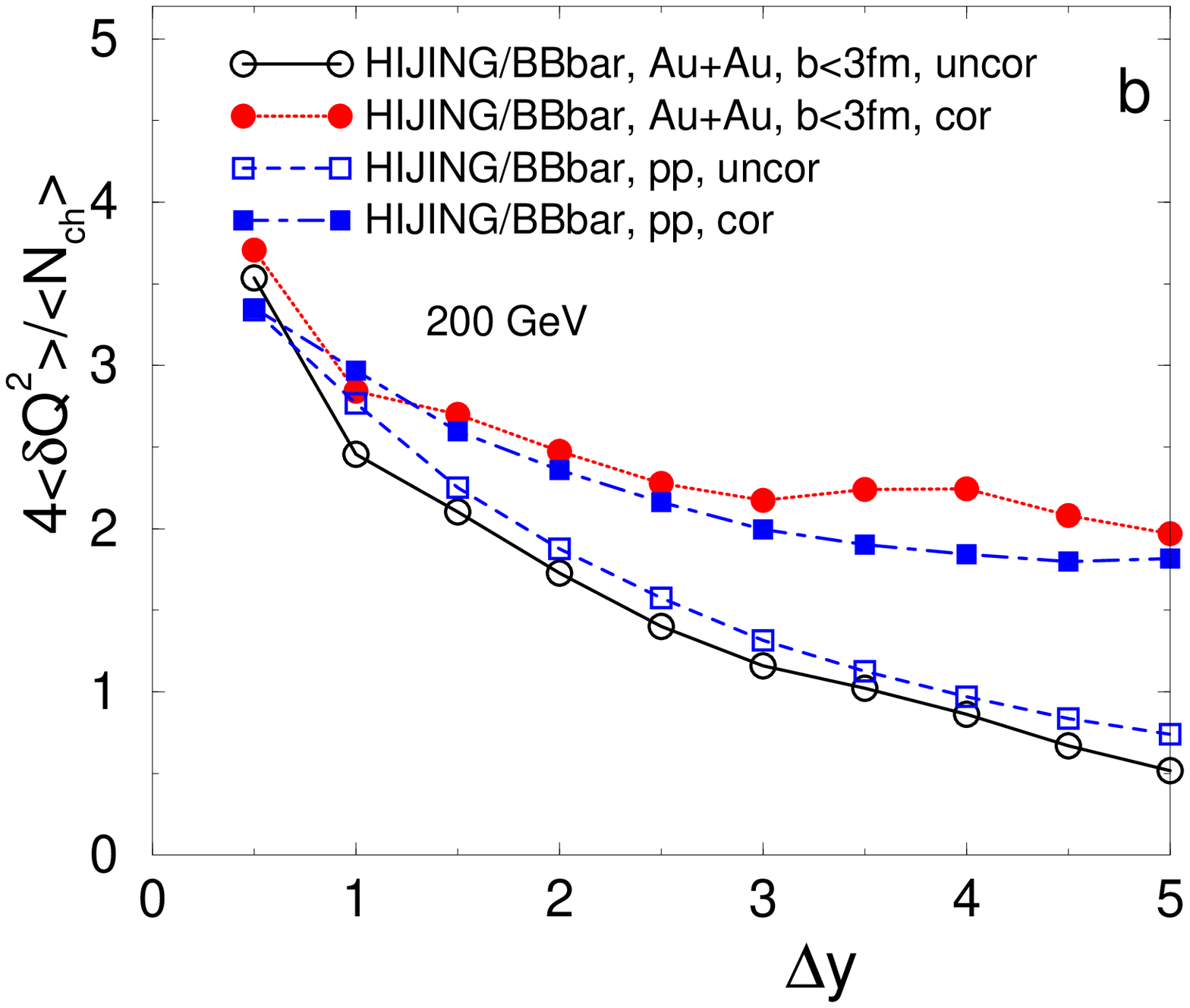}}
\end{figure}
\begin{figure}[ht]\epsfxsize=7.cm
\centerline{\epsfbox{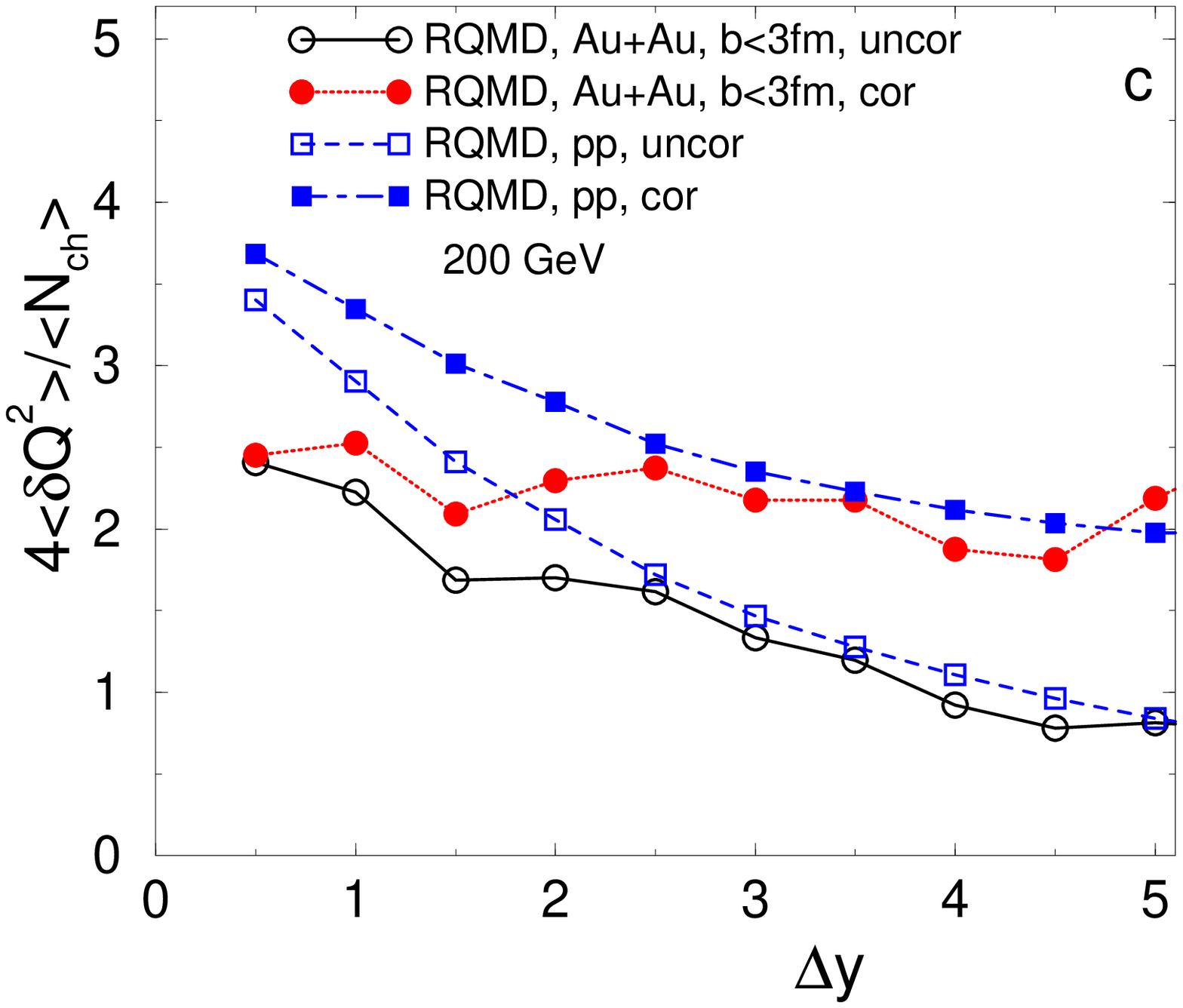}}
\end{figure}
\begin{figure}[ht]\epsfxsize=7.cm
\centerline{\epsfbox{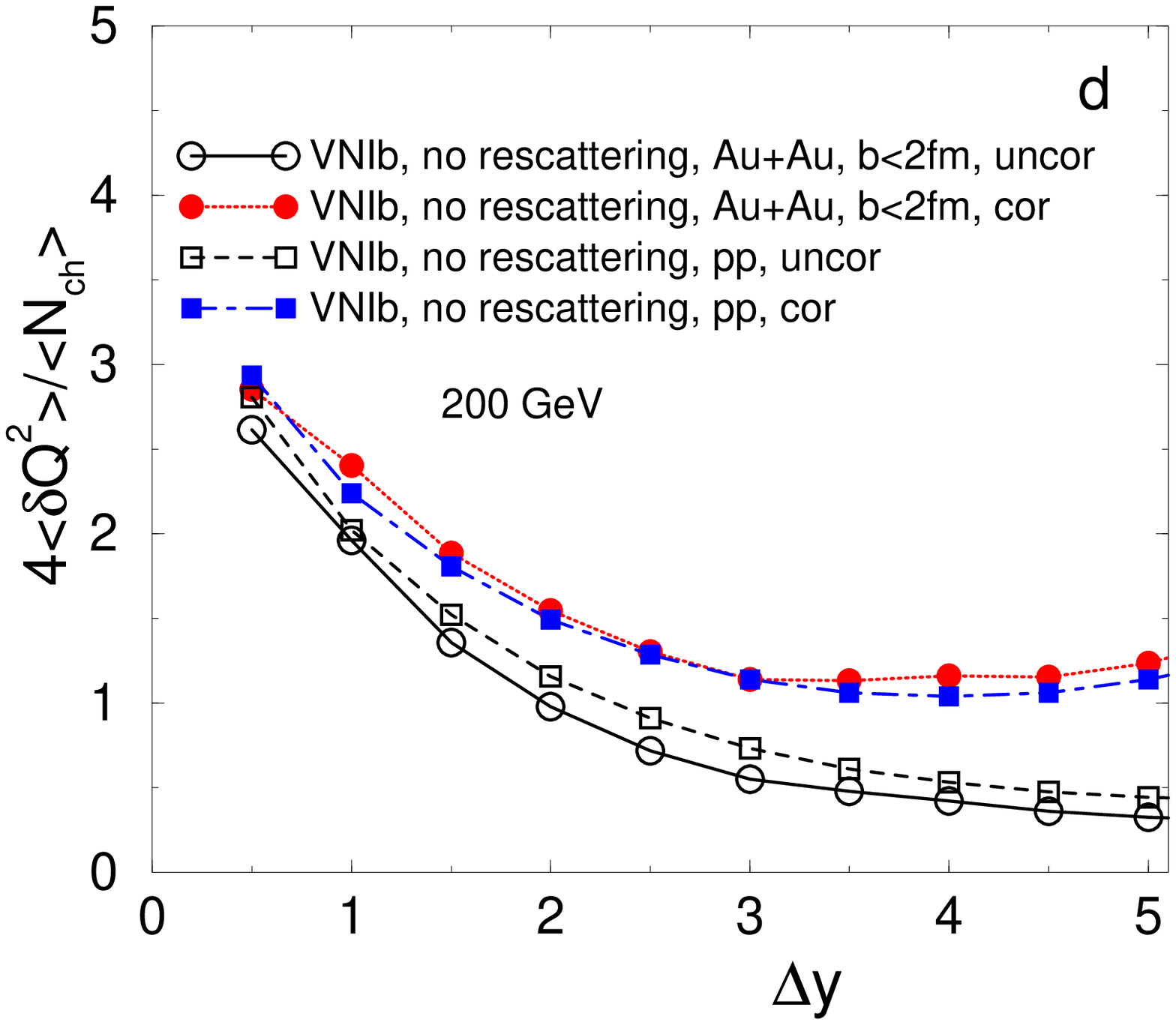}}
\end{figure}
\begin{figure}[ht]\epsfxsize=7.cm
\centerline{\epsfbox{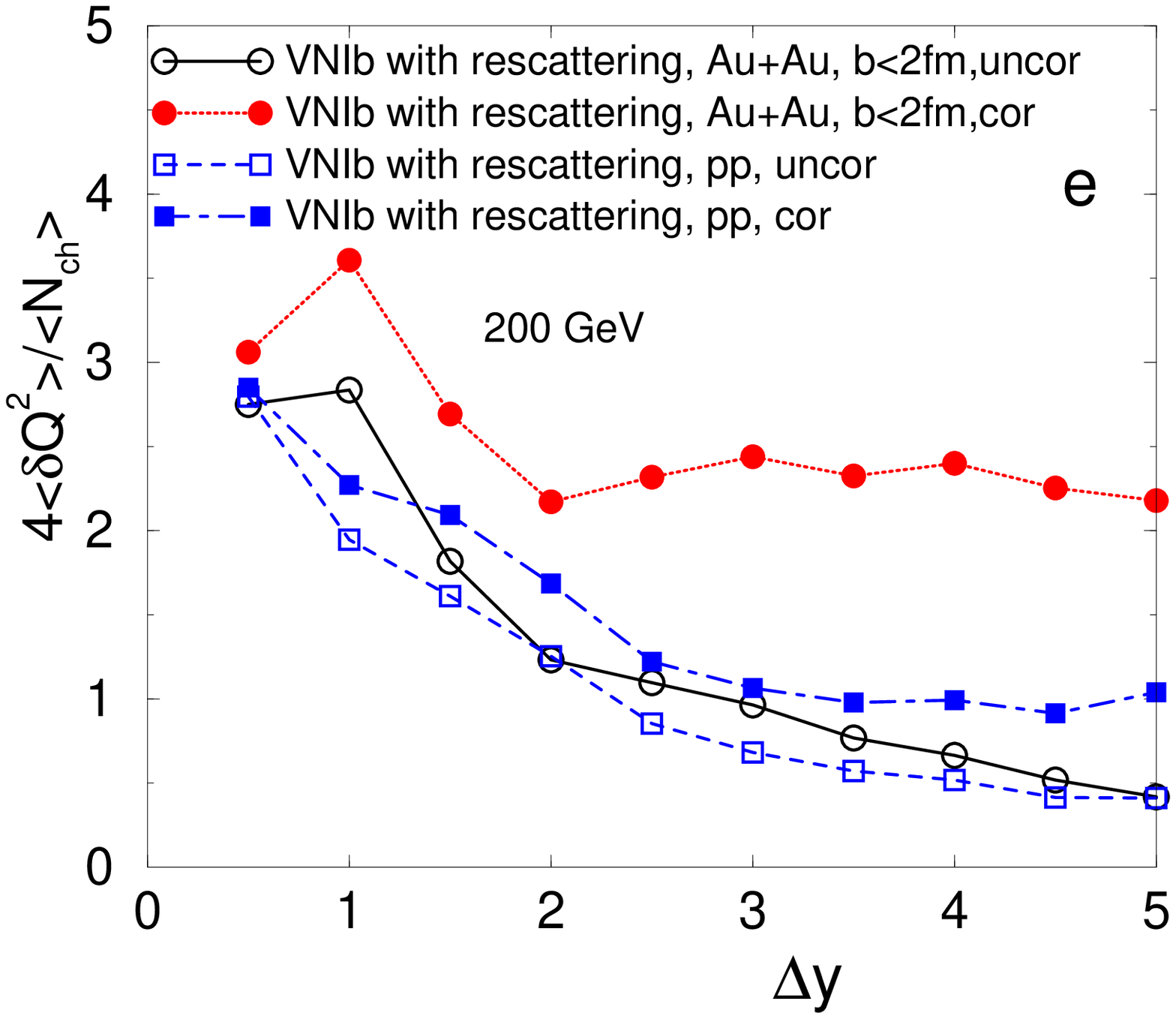}}
\caption{$D$-measure values from  
(a) HIJING v1.35; (b) HIJING/$B{\bar B}$ v1.10;
(c) RQMD v2.4 (with rescattering); (d) VNIb (rescattering turned off);
(e) VNIb (rescattering turned on) models vs. rapidity 
$y_{cm} \pm \frac{\Delta y}{2}$ for Au+Au central collisions 
(full symbols) and and pp collisions (empty symbols) at total nucleon-nucleon c.m. energy
$\sqrt{s}_{NN}$=200 GeV.}
\label{fig12}
\end{figure}

\begin{figure}[ht]\epsfxsize=8cm
\centerline{\epsfbox{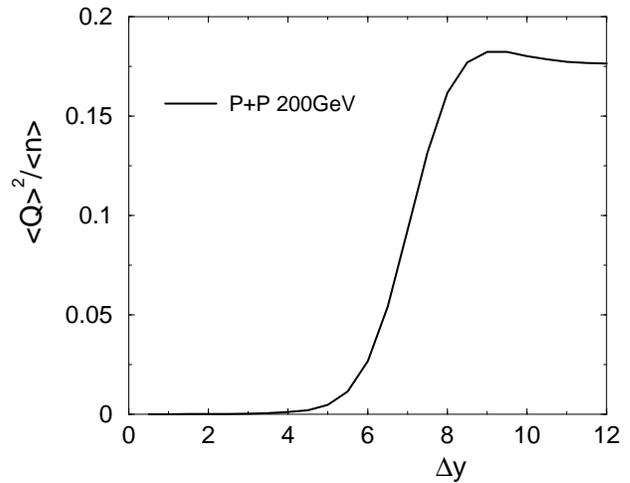}}
\caption{$\frac{\langle Q \rangle^2}{\langle N_{ch}\rangle}$ vs. 
 rapidity  $y_{cm} \pm \frac{\Delta y}{2}$ for pp  collisions 
at total c.m. energy $\sqrt{s}$=200 GeV.}
\label{fig9}
\end{figure}

The interesting result is that 
$D$ values are similar for 
$pp$, and $Au+Au$ collisions for all MCGs when rescattering effects are  
neglected.  
In the following, we try to explain this in the framework of a ``participant model''\cite{BH99}.
As in Ref.\cite{BH99}, we write 
\begin{equation}
Q=\sum_{i=1}^{N_p} Q_i   .
\end{equation}
Here $Q$ is the total charge of $AA$ collisions, $Q_i$ is the 
charge produced by each nucleon +nucleon ($n+n$) collisions in a 
specific rapidity window and $N_p$ is the number of 
$nn$ collisions for each $AA$ collisions. Taking the average over a 
number of events we have 
\begin{equation}
\langle Q \rangle =\langle N_p\rangle \langle Q_i\rangle
\label{e29}
\end{equation}
and 
\begin{equation}
\langle Q^2 \rangle =\langle N_p\rangle \langle Q_i^2\rangle
+\langle N_p(N_p-1)\rangle \langle Q_i\rangle^2.
\end{equation}
In the derivation we have used 
$\langle Q_i Q_j\rangle=\langle Q_i\rangle \langle Q_j\rangle$. The mean charged multiplicity 
for $AA$ collisions can be expressed as\cite{BH99} 
\begin{equation}
\langle N_{ch}\rangle =\langle N_p\rangle \cdot \langle n_i\rangle.
\end{equation} 
Here $n_i$ is the charged particles produced by each $n+n$ collisions. 
Finally we get the following equation:
\begin{equation}
\frac{\langle \delta Q^2\rangle}{\langle N_{ch}\rangle}=
\frac{\langle \delta Q_i^2\rangle}{\langle n_i\rangle}+\frac{\langle Q_i\rangle^2}{\langle n_i\rangle}
\frac{\langle \delta N_p^2\rangle}{\langle N_p\rangle}   .
\end{equation}
If the $P(N_p)$ distribution is Poissonian, then 
$ \frac{\langle \delta N_p^2\rangle}{\langle N_p\rangle}=1$. In Ref.\cite{BH99}, the author estimated 
the above value to be around $1.1$. 
$\frac{\langle Q_i\rangle^2}{\langle n_i\rangle}$ should be much less than one for very high energy 
$n+n$ collisions, that is 
\begin{equation}
\langle n_i\rangle >>\langle Q_i\rangle.
\end{equation}
To confirm this, we plot the ratios of $\frac{\langle Q_i\rangle^2}{\langle n_i\rangle}$ vs. rapidity 
window for $pp$ collisions at $200\,{\em GeV}$ in Fig. \ref{fig9}. One sees clearly, that for smaller rapidity windows  
the ratio is near zero,
while for whole window the value is around 0.2. The later is due to 
charge conservation effects. For larger rapidity regions, 
the particles are produced mainly near the leading 
valence quarks, so we notice that there is a sharp increase of the value 
$\frac{\langle Q\rangle^2}{\langle N_{ch}\rangle}$ for larger rapidity 
window. For the inner part of the rapidity region, due to the charge 
conservation, the mean charge $\langle Q_i\rangle \sim 0$.  From above 
figures we can safely 
say that $D$-measure for $AA$ collisions should be roughly the same as 
for $pp$ case when the rescattering effects are negligible.  
Any deviation between the $D$-measure of $AA$ 
and $pp$ may indicate a signature of new physics 
in $AA$ collisions.  Thus, it is necessary to check the consistency 
between $pp$ 
collisions and $AA$ collisions
results before we may conclude that $D$-measure is a signature of QGP. 

\section{Comments and conclusions}
 
Theoretical predictions of $D$-measure from 
VNIb, HIJING v1.35, HIJING/B${\bar B}$ v1.10, 
RQMD v2.4, indicate that 
the fluctuation of charge is sensitive to the 
parton number embedded in the model 
if the rescattering effects are not essential; therefore, 
$D$-measure can be a signature of QGP. 

However, if the charge fluctuation shows  no  
impact parameter dependence, then we have to slightly change our 
views. 
If we observe similar signal for pp and for peripheral collisions,
the charge fluctuation could be only a signature of the fundamental 
degrees of freedom that we need to take into consideration in the collision 
processes. 
In other words, charge fluctuation 
can tell us when we should treat the heavy-ion collisions 
as simple hadronic cascade or when is necessary to use 
QCD, or some model in between. 
This idea has been 
used in $e^+e^-$ collisions to see when one should use a  
cluster picture and when one needs to use a parton picture\cite{Ochs}.
If the $D$-measure for $AA$ is bigger than 
the $D$-measure for $pp$, there could exist a stronger rescattering 
effects in heavy-ion collisions; on the other hand, if the $D$-measure for 
$AA$ is lower than the $D$-measure for $pp$, 
some new physics in $AA$ collisions should be involved. 
Upcoming experiments at RHIC 
will allow us to draw more definite conclusions.

To consider $D$-measure as a signature of QGP, one must certify that 
the model from Ref.\cite{JK00} and Ref.\cite{AHM00} can not be 
applied for pp collisions. If a single thermal model 
is valid for  both $AA$ and $pp$ collisions, then 
the conclusions that $D$-measure is a signature of QGP 
may be questionable. 
 If  the statistical model for parton degrees of freedom 
 can also be used in pp collisions  we can not 
see any reason why the predictions from pp collisions 
should be different from the prediction of $AA$\cite{comment}, but we totally 
agree that the $D$-measure can tell us if we need to 
consider partonic or hadronic degrees of freedom in the collisions.  

Our theoretical predictions using different MCG models show
that $D$-measure is sensitive to different 
parton content embedded in the model if the rescattering effects 
is not dominant. 
We find that the $D$-measure values do not depend on impact 
parameter for RQMD v2.4 model and also 
we obtain similar results for AA and pp collisions,
and we explain this using the participant model. 
On the other hand, we find that 
the values of $D$-measure from  VNIb model are strongly
dependent on rescattering effects which spoil the original 
signature from the initial state of collisions. 
However, any deviation among the prediction of $D$-measure for different 
impact parameter in AA collisions and pp collisions may indicate  
that the rescattering effects play a key role in interactions, or 
a signature for new physics (e.g. presumably QGP) in AA collisions.
Note that a recent paper\cite{SS00} was concerned about the specific 
effects of rescattering on the $D$-measure. Within the framework of 
existing empirical models, our work can be seen as a 
quantitative answer to those questions. Also it will be crucial to 
repeat the calculations done here with the soon-to-be-release 
next version of the parton cascade code\cite{BS}.

Recently, the STAR collaboration  has analyzed the $D$-measure at RHIC energy 
($\sqrt{s_{NN}}=130\,\, $GeV) and 
has found that the $D$-measure value is around three and 
has no centrality dependence\cite{Star}. 
This results are consistent with our 
prediction and those of Ref.\cite{BJK00}. 
In the calculation of the STAR collaboration\cite{Star}, they 
did not use the correction which accounts for the net 
charge and global charge conservation; if we consider 
this correction, the $D$-measure will be around $3.9$. 
 However, the high value of $D$-measure 
does not imply that QGP is not formed at RHIC, this high 
value of $D$-measure may still be 
explained by final state rescattering. 
%
%

\begin{center}
{\bf Acknowledgment}
\end{center}

This work was partly supported by the 
Natural Science and Engineering Research Council of Canada and 
the Fonds FCAR of the Quebec Government. The authors would 
like to thank U. Heinz, V. Koch, S. Mrowczynski,
S. Voloshin and R. Wit for helpful communications.


\begin{thebibliography}{29}
\bibitem{Hijing}
In HIJING, the authors also include a gluon contribution
( X. N. Wang, M. Gyulassy, nucl-th/0008014).  
 However, the gluon density is not so high, compared with VNIb; therefore, 
gluon effects on the charged fluctuation are not important.
\bibitem{comment}
We mean here that the parton distribution is the same for both pp and AA.
If there are some differences between the 
partons distribution in pp and AA, then we may have different 
results for pp and AA.
\end{thebibliography}
\end {document}